# Analytical Investigation of Magneto-Plasmons in Anisotropic Multi-layer Planar Waveguides Incorporating Magnetically Biased Graphene Sheets

Mohammad Bagher Heydari [1,*], Mohammad Hashem Vadjed Samiei [1]

[1,*] School of Electrical Engineering, Iran University of Science and Technology (IUST), Tehran, Iran

[*]Corresponding author: mo_heydari@alumni.iust.ac.ir ; heydari.sharif@gmail.com

**Abstract** This article proposes a novel analytical model for the anisotropic multi-layer structures containing magnetically biased graphene sheets. The multi-layer structure is composed of various magnetic materials, each one has the permittivity and permeability tensors of $\bar{\bar{\varepsilon}}$ and $\bar{\bar{\mu}}$, respectively. An external magnetic field is applied, normal to the structure surface. Each graphene sheet, with anisotropic conductivity tensor ($\bar{\bar{\sigma}}$), has been sandwiched between two adjacent magnetic materials. Our model is used to find the dispersion relation of the structure. Now, obtaining plasmonic features of the structure, such as the effective index and propagation loss is straightforward. Four exemplary structural variants have been investigated to show the richness of the proposed general structure regarding the related specific plasmonic wave phenomena and effects. These aspects are essential to form our structural design platform to propose novel plasmonic devices such as biosensors, modulators, absorbers, transparent electrodes and tunable metamaterials in THz frequencies. A very good agreement between the analytical and full-wave simulation results is seen.

## 1. Introduction

Graphene is one of the interesting two-dimensional materials which offers a number of fundamentally fascinating features. These properties make graphene a good candidate for designing of novel photonic and electronic devices such as waveguides [1-13], circulator [14, 15], coupler [16], resonator [17], and filter [18]. Graphene plasmonics is a new research area emerged in recent years, which has been developed based on the optical conductivity of graphene that can be varied by either electrostatic or magnetostatic gating. In near-infrared frequencies, metal plasmonics is a very promising candidate for various applications [19, 20].

The first studies concerned the propagation of surface plasmon polaritons (SPPs) on the graphene sheet was done by George Hanson [21-23]. He studied the interaction of the current source with an anisotropic graphene sheet placed between two dielectric layers in [23]. In [21], the dyadic Green's function for the infinite graphene sheet was obtained, similar to [23]. Afterward, the number of research articles on graphene-based planar waveguides was increased and the researchers focused on planar structures due to their simplicity and easy-to-design procedures [24-31]. For instance, Giampiero Lovat proposed a transverse equivalent network for graphene-based waveguide, where a graphene layer had been deposited on the top of the grounded dielectric [25]. In [27], Diego Correas-Serrano et al. studied graphene single and parallel plate waveguides and offered a circuit model, which could calculate the modal properties accurately.

Between the graphene-based structures, the study on plasmonic characteristics of multi-layered structures has attracted the attention of many researchers [22, 32-38]. The first survey on plasmonic modes of three and five-layer parallel-plate waveguides formed by graphene sheets was done in [22]. It was shown that these waveguides could support quasi-TEM modes with attenuation factors similar to metallic plasmonic waveguides [22]. In [32], the spectral-domain representations were achieved for Green's functions of multi-layer graphene-based structures by using a propagator matrix technique. It should be noted that most of the research papers in this field have focused on calculating the transmittance, the absorption and the reflection coefficients of the electromagnetic waves when a normally incident wave illuminates the multi-layered structure [35-37]. For instance, a transfer matrix method has been proposed to calculate the reflectance and the transmittance for TE and TM polarizations in [37]. In this work [37], the anisotropic graphene sheet was located between two dielectric layers in the multi-layered structure, which could excite hybrid TE-TM plasmons along with the graphene sheet. Recently, multi-layered structures with graphene sheets have been utilized in many interesting applications such as tunable metamaterials [39-41], absorbers [42, 43], biosensors [44], modulators [45], transparent electrodes [46, 47], Faraday rotation based structures [48, 49], and second-harmonic generation [50]. Due to these interesting applications, a novel analytical model has been proposed in this article for studying of tunable magneto-plasmons in the general anisotropic multi-layered structures with magnetically biased graphene sheets.

We present a complete modal analysis of the structure leading to closed-form relations of field distributions of all propagating modes. Based on the uniqueness theorem, every other solution obtained by arbitrary other methods should be equivalent to our result. Regarding the speed of calculation, there is no numerical method faster than an analytical solution based on closed-form relations, considering the fact that numerical methods are volumetric and need discretizing of the structure. In summary, our analytical model has two fascinating advantages: first, it is able to calculate the propagating parameters of the general structure very quickly and second, it has high accuracy compared to other computational methods. To the authors' knowledge, no published work has been presented the analytical study of the general anisotropic multi-layered structures with magnetically-biased graphene layers. In our proposed structure, each layer has the permittivity and permeability tensors of $\bar{\bar{\varepsilon}}$ and $\bar{\bar{\mu}}$, respectively and the anisotropic graphene sheet has been placed between the two adjacent anisotropic materials. An external magnetic field is applied normally to the structure surface. This complex configuration allows one to adjust and control the plasmonic properties of the structure by varying the chemical doping of graphene, the electrostatic magnetic bias, altering the number of layers, and choosing any type of material (such as magnetic material, ferrite, dielectric, gyro-electric) for each layer. The magnetic materials have some specific features such as the possibility of generating plasma waves, and the excitation of helicons (when the external magnetic field is applied) [51-55]. Hence, our tunable structure can be utilized to design new plasmonic devices such as tunable metamaterials, modulators, high efficient absorbers and transparent electrodes at THz frequencies.

The paper is organized as follows. In section 2, general equations for the bi-gyrotropic media will be briefly summarized by using Maxwell's equations. Then, section 3 presents the analytical model for the multi-layered structure with anisotropic graphene sheets by utilizing the mathematical relations outlined in section 2. This section expresses the derivation of the model by applying the mode-matching technique. Section 4 studies four exemplary structural variants. The first and the second structures have been chosen the simple, familiar graphene-based waveguides to verify the performance and the high accuracy of the theoretical formulations outlined in section 3. The first example investigates a graphene sheet, with an anisotropic conductivity tensor, sandwiched between two dielectric layers. The graphene layer has been biased magnetically in this structure, which supports hybrid TE-TM plasmonic mode. The results show that a very high effective index of about 3900 at the frequency 32 THz is achievable for the external magnetic bias of B= 4T. The second example is a three-layer structure, constituting Air-Graphene-$SiO_2$-Si layers. In this structure, the graphene layer has been deposited on the $SiO_2$-Si layer and its conductivity is characterized by isotropic surface conductivity ($\sigma$).

To show the richness of the proposed general structure regarding the related specific plasmonic wave phenomena and effects, two novel graphene-based waveguides have been suggested and investigated as the third and the fourth structures. The third example introduces a new ferrite-graphene waveguide, which has a non-reciprocity effect. This hybrid waveguide is able to control the propagating properties of plasmonic waves both with chemical potential and magnetic bias. It should be mentioned that the usage of the ferrite materials at low THz frequencies has been reported in some research articles [56-61]. One of the drawbacks of utilizing the ferrite materials at the THz region is their necessity for high external magnetic fields to change the permeability at this region [56, 62]. However, some novel ferrite materials have been studied and introduced at the THz frequencies, which require small external magnetic fields [62-64]. Therefore, it seems that ferrite is a good candidate for designing tunable devices in low THz frequencies. As the fourth example, a planar graphene-based structure with a gyro-electric substrate has been introduced and considered. The gyroelectric layer has an anisotropic permittivity, which produces the non-reciprocity effect. In this tunable structure, the plasmonic features can be varied via the chemical doping and external magnetic field, similar to the third waveguide. To authors' knowledge, the hybridization of magnetically biased graphene sheet with the ferrite and the gyro-electric materials, for an external magnetic field applied normally to the structure surface, have been reported in this article for the first time. Finally, section 5 concludes the article.

## 2. General Equations in Bi-gyrotropic Media

This section studies the governing equations inside the bi-gyrotropic media. These mathematical relations will be utilized in the next section to obtain the accurate and novel analytical model of our proposed structure. First, consider Maxwell's equations in the frequency domain (suppose $e^{j\omega t}$) [65]:

$$\nabla \times \mathbf{E} = -j\omega \bar{\bar{\mu}} \cdot \mathbf{H} \tag{1}$$

$$\nabla \times \mathbf{H} = j\omega \bar{\bar{\varepsilon}} \cdot \mathbf{E} \tag{2}$$

Where $\bar{\bar{\varepsilon}}$ and $\bar{\bar{\mu}}$ are the permittivity and permeability tensors of bi-gyrotropic material, respectively. In the presence of the external magnetic field in the z-axis, these tensors are described for magnetic materials as follows [65]:

$$\bar{\bar{\mu}} = \mu_0 \begin{pmatrix} \mu & j\mu_a & 0 \\ -j\mu_a & \mu & 0 \\ 0 & 0 & \mu_\parallel \end{pmatrix} \tag{3}$$

$$\bar{\bar{\varepsilon}} = \varepsilon_0 \begin{pmatrix} \varepsilon & j\varepsilon_a & 0 \\ -j\varepsilon_a & \varepsilon & 0 \\ 0 & 0 & \varepsilon_\parallel \end{pmatrix} \tag{4}$$

Where $\varepsilon_0$ and $\mu_0$ are the permittivity and permeability of the free space, respectively. In (3), the diagonal and off-diagonal elements are expressed as [65]:

$$\mu = 1 + \frac{\omega_M (\omega_H + j\omega\alpha)}{\omega_H^2 - (1+\alpha^2)\omega^2 + 2j\alpha\omega\omega_H} \tag{5}$$

$$\mu_a = \frac{\omega \omega_M}{\omega_H^2 - (1+\alpha^2)\omega^2 + 2j\alpha\omega\omega_H} \tag{6}$$

$$\mu_\parallel = 1 - \frac{j\alpha \omega_M}{\omega + j\alpha \omega_H} \tag{7}$$

In the above relations, $\omega_H = \gamma H_0$ and $\omega_M = \gamma M_S$, $M_S$ is the saturation magnetization, $\gamma$ is the gyromagnetic ratio, and $\alpha$ is the Gilbert damping constant. The diagonal and off-diagonal elements of (4) have familiar relations [66]:

$$\varepsilon = \varepsilon_\infty \left(1 - \frac{\omega_p^2 (\omega + j\upsilon)}{\omega\left[(\omega + j\upsilon)^2 - \omega_c^2\right]}\right) \tag{8}$$

$$\varepsilon_a = \varepsilon_\infty \left(\frac{\omega_p^2 \omega_c}{\omega\left[(\omega + j\upsilon)^2 - \omega_c^2\right]}\right) \tag{9}$$

$$\varepsilon_{\|} = \varepsilon_{\infty}\left(1 - \frac{\omega_p^2}{\omega(\omega + j\upsilon)}\right) \tag{10}$$

In (8)-(10), $\nu$ is the effective collision rate and $\varepsilon_{\infty}$ is the background permittivity. We have determined the permittivity tensor assuming a specific value of the collision rate ($\nu$). It is worth to be mentioned that the effective collision rate in gyro-electric materials, such as n-type InSb, depends on the mobility of the charge carriers [67, 68]. The mobility and carrier concentration change as the temperature varies. Hence, the effective collision rate depends on the charge concentration ($n_s$) [67-70]. The plasma and the cyclotron frequencies are defined as [66]:

$$\omega_p = \sqrt{\frac{n_s e^2}{\varepsilon_0 \varepsilon_{\infty} m^*}} \tag{11}$$

$$\omega_c = \frac{e B_0}{m^*} \tag{12}$$

Where $e, m^*$ and $n_s$ are the charge, effective mass and the density of the carriers. From Maxwell's equations in cylindrical coordinates, the z-components of the electric and magnetic fields inside the bi-gyrotropic layer satisfy [65]:

$$\left(\nabla_{\perp}^2 + \frac{\varepsilon_{\|}}{\varepsilon}\frac{\partial^2}{\partial z^2} + (k_0^2 \varepsilon_{\|} \mu_{\perp})\right) E_z + k_0 \mu_{\|}\left(\frac{\varepsilon_a}{\varepsilon} + \frac{\mu_{\alpha}}{\mu}\right)\frac{\partial}{\partial z} H_z = 0 \tag{13}$$

$$\left(\nabla_{\perp}^2 + \frac{\mu_{\|}}{\mu}\frac{\partial^2}{\partial z^2} + (k_0^2 \varepsilon_{\perp} \mu_{\|})\right) H_z - k_0 \varepsilon_{\|}\left(\frac{\varepsilon_a}{\varepsilon} + \frac{\mu_{\alpha}}{\mu}\right)\frac{\partial}{\partial z} E_z = 0 \tag{14}$$

where

$$\nabla_{\perp}^2 = \frac{1}{r}\frac{\partial}{\partial r} r \frac{\partial}{\partial r} + \frac{1}{r^2}\frac{\partial}{\partial^2 \varphi} \tag{15}$$

In these equations, $k_0$ is the free space wave-number and,

$$\varepsilon_{\perp} = \varepsilon - \frac{\varepsilon_{\alpha}^2}{\varepsilon} \tag{16}$$

$$\mu_{\perp} = \mu - \frac{\mu_{\alpha}^2}{\mu} \tag{17}$$

To find the radial modes in the bi-gyrotropic media, the z-components of the electric and magnetic fields are written

$$H_z(r,\varphi,z) = \int_{-\infty}^{+\infty}\sum_{m=-\infty}^{\infty} H_m(z)\exp(-jm\varphi) J_m(k_{\rho}r) dk_{\rho} \tag{18}$$

$$E_z(r,\varphi,z) = \int_{-\infty}^{+\infty}\sum_{m=-\infty}^{\infty} E_m(z)\exp(-jm\varphi) J_m(k_{\rho}r) dk_{\rho} \tag{19}$$

In the above equations, $m$ is an integer and $k_{\rho}$ is the propagation constant of radial waves. Now, by substituting (18) and (19) into (13) and (14), one can obtain the following coupled equations:

$$\frac{\varepsilon_{\|}}{\varepsilon}\frac{\partial^2 E_m(z)}{\partial z^2} + \left(k_0^2 \varepsilon_{\|}\mu_{\perp} - k_{\rho}^2\right) E_m(z) + k_0 \mu_{\|}\left(\frac{\varepsilon_a}{\varepsilon} + \frac{\mu_{\alpha}}{\mu}\right)\frac{\partial H_m(z)}{\partial z} = 0 \tag{20}$$

$$\frac{\mu_{\|}}{\mu}\frac{\partial^2 H_m(z)}{\partial z^2} + \left(k_0^2 \varepsilon_{\perp}\mu_{\|} - k_{\rho}^2\right) H_m(z) - k_0 \varepsilon_{\|}\left(\frac{\varepsilon_a}{\varepsilon} + \frac{\mu_{\alpha}}{\mu}\right)\frac{\partial E_m(z)}{\partial z} = 0 \tag{21}$$

These equations are combined and then a fourth-order differential equation is achieved

$$\frac{\partial^4 H_m(z)}{\partial z^4} + \left(\frac{\mu\varepsilon}{\mu_\| \varepsilon_\|}\right)\left[\left(k_0^2 \varepsilon_\| \mu_\perp - k_\rho^2\right)\frac{\mu_\|}{\mu} + \left(k_0^2 \varepsilon_\perp \mu_\| - k_\rho^2\right)\frac{\varepsilon_\|}{\varepsilon} + k_0^2 \mu_\| \varepsilon_\| \left(\frac{\varepsilon_a}{\varepsilon} + \frac{\mu_\alpha}{\mu}\right)^2\right]\frac{\partial^2 H_m(z)}{\partial z^2} + \tag{22}$$

$$\left(\frac{\mu\varepsilon}{\mu_\| \varepsilon_\|}\right)\left[\left(k_0^2 \varepsilon_\| \mu_\perp - k_\rho^2\right)\cdot\left(k_0^2 \varepsilon_\perp \mu_\| - k_\rho^2\right)\right] H_m(z) = 0$$

By defining the following coefficients,

$$A_1 = \left(\frac{\mu\varepsilon}{\mu_\| \varepsilon_\|}\right)\left(\left(k_0^2 \varepsilon_\| \mu_\perp - k_\rho^2\right)\frac{\mu_\|}{\mu} + \left(k_0^2 \varepsilon_\perp \mu_\| - k_\rho^2\right)\frac{\varepsilon_\|}{\varepsilon} + k_0^2 \mu_\| \varepsilon_\| \left(\frac{\varepsilon_a}{\varepsilon} + \frac{\mu_\alpha}{\mu}\right)^2\right) \tag{23}$$

$$A_2 = \left(\frac{\mu\varepsilon}{\mu_\| \varepsilon_\|}\right)\left[\left(k_0^2 \varepsilon_\| \mu_\perp - k_\rho^2\right)\cdot\left(k_0^2 \varepsilon_\perp \mu_\| - k_\rho^2\right)\right] \tag{24}$$

The characteristics equation of (22) is

$$s^4 + A_1 s^2 + A_2 = 0 \tag{25}$$

In general form, the solutions of (25) are written as

$$H_m(z) = A_m^+ e^{k_{z,1} z} + B_m^+ e^{k_{z,2} z} + A_m^- e^{-k_{z,1} z} + B_m^- e^{-k_{z,2} z} \tag{26}$$

In (26), $k_{z,1}, k_{z,2}$ are

$$k_{z,1} = \sqrt{\frac{-A_1 + \sqrt{A_1^2 - 4A_2}}{2}} \tag{27}$$

$$k_{z,2} = \sqrt{\frac{-A_1 - \sqrt{A_1^2 - 4A_2}}{2}} \tag{28}$$

To obtain the electric field $E_m(z)$, we should substitute (26) into (21), which yields to

$$E_m(z) = T_1^+ A_m^+ e^{+k_{z,1} z} + T_2^+ B_m^+ e^{+k_{z,2} z} + T_1^- A_m^- e^{-k_{z,1} z} + T_2^- B_m^- e^{-k_{z,2} z} \tag{29}$$

Where

$$T_i^\pm = \frac{\pm 1}{k_0 \varepsilon_\| \left(\frac{\varepsilon_a}{\varepsilon} + \frac{\mu_\alpha}{\mu}\right)}\left(\frac{\mu_\|}{\mu} k_{z,i} + \frac{1}{k_{z,i}}\left(k_0^2 \varepsilon_\perp \mu_\| - k_\rho^2\right)\right) \quad i = 1, 2 \tag{30}$$

It should be noted that $i$-index in relation (30) shows the roots of the characteristics equation of (25) and the sign $\pm$ indicates the exponential form. For instance, $T_2^-$ is related to the coefficient of $\exp(-k_{z,2} z)$. Finally, the transverse components of electric and magnetic fields are expressed:

$$\begin{pmatrix} E_{r,i}^\pm \\ H_{r,i}^\pm \end{pmatrix} = \bar{\bar{Q}}_i^{Pos,\pm} \frac{\partial}{\partial r}\begin{pmatrix} E_{z,i}^\pm \\ H_{z,i}^\pm \end{pmatrix} + \frac{m}{r}\bar{\bar{Q}}_i^{Neg,\pm}\begin{pmatrix} E_{z,i}^\pm \\ H_{z,i}^\pm \end{pmatrix} \quad i = 1, 2 \tag{31}$$

$$j\begin{pmatrix} E_{\varphi,i}^\pm \\ H_{\varphi,i}^\pm \end{pmatrix} = \bar{\bar{Q}}_i^{Neg,\pm} \frac{\partial}{\partial r}\begin{pmatrix} E_{z,i}^\pm \\ H_{z,i}^\pm \end{pmatrix} + \frac{m}{r}\bar{\bar{Q}}_i^{Pos,\pm}\begin{pmatrix} E_{z,i}^\pm \\ H_{z,i}^\pm \end{pmatrix} \quad i = 1, 2 \tag{32}$$

In the above relations, Q-matrices are defined:

$$\bar{\bar{Q}}_i^{Pos,\pm} = \frac{1}{2}\left[\frac{1}{k_{z,i}^2 + k_0^2\varepsilon_+\mu_+}\begin{pmatrix} \pm k_{z,i} & -\omega\mu_0\mu_+ \\ \omega\varepsilon_0\varepsilon_+ & \pm k_{z,i} \end{pmatrix} + \frac{1}{k_{z,i}^2 + k_0^2\varepsilon_-\mu_-}\begin{pmatrix} \pm k_{z,i} & \omega\mu_0\mu_- \\ -\omega\varepsilon_0\varepsilon_- & \pm k_{z,i} \end{pmatrix}\right] \quad (33)$$

$$\bar{\bar{Q}}_i^{Neg,\pm} = \frac{1}{2}\left[\frac{1}{k_{z,i}^2 + k_0^2\varepsilon_+\mu_+}\begin{pmatrix} \pm k_{z,i} & -\omega\mu_0\mu_+ \\ \omega\varepsilon_0\varepsilon_+ & \pm k_{z,i} \end{pmatrix} - \frac{1}{k_{z,i}^2 + k_0^2\varepsilon_-\mu_-}\begin{pmatrix} \pm k_{z,i} & \omega\mu_0\mu_- \\ -\omega\varepsilon_0\varepsilon_- & \pm k_{z,i} \end{pmatrix}\right] \quad (34)$$

and

$$\mu_\pm = \mu \pm \mu_a \quad (35)$$

$$\varepsilon_\pm = \varepsilon \pm \varepsilon_a \quad (36)$$

## 3. The Proposed General Structure and Formulation of the Problem

In the previous section, the general relations of bi-gyrotropic media were studied in detail. Here, we will utilize the mathematical relations outlined in the previous section to find the dispersion relation of the general multi-layer structure. In the next step, obtaining plasmonic features of the structure, such as the effective index and propagation loss is straightforward. Fig.1 represents the schematic of the general multi-layered structure, where each magnetic material has the permittivity and permeability tensors of $\bar{\bar{\varepsilon}}, \bar{\bar{\mu}}$ and the anisotropic graphene sheet ($\bar{\bar{\sigma}}$) has been located between two adjacent anisotropic layers. The electric and magnetic currents, which have been considered as the excitation sources, have been located at the top of the whole structure. The structure has been magnetized along the z-direction by a DC bias magnetic field $B_0$.

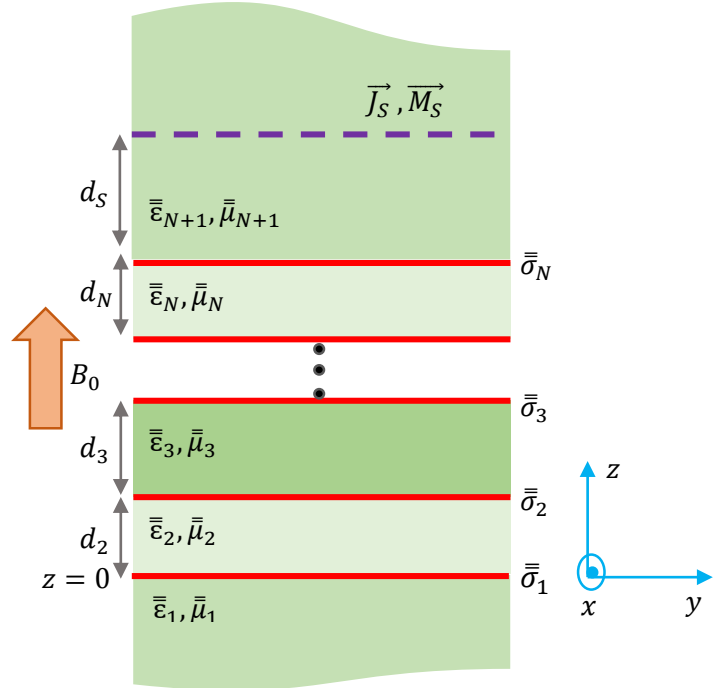

**Fig. 1.** General multi-layered structure with magnetic materials and anisotropic graphene sheets. The electric and magnetic surface currents have been located at the top of the whole waveguide.

Let us suppose that the *N*-th layer of the bi-gyrotropic layer has the following permeability and permittivity tensors:

$$\bar{\bar{\varepsilon}}_N = \varepsilon_0 \begin{pmatrix} \varepsilon_N & j\varepsilon_{a,N} & 0 \\ -j\varepsilon_{a,N} & \varepsilon_N & 0 \\ 0 & 0 & \varepsilon_{\|,N} \end{pmatrix} \tag{37}$$

$$\bar{\bar{\mu}}_N = \mu_0 \begin{pmatrix} \mu_N & j\mu_{a,N} & 0 \\ -j\mu_{a,N} & \mu_N & 0 \\ 0 & 0 & \mu_{\|,N} \end{pmatrix} \tag{38}$$

The diagonal and off-diagonal elements for this general material are defined similarly to the relations (5)-(10). Moreover,

$$\varepsilon_{\perp,N} = \varepsilon_N - \frac{\varepsilon_{a,N}^2}{\varepsilon_N} \tag{39}$$

$$\mu_{\perp,N} = \mu_N - \frac{\mu_{a,N}^2}{\mu_N} \tag{40}$$

$$\varepsilon_{\pm,N} = \varepsilon_N \pm \varepsilon_{a,N} \tag{41}$$

$$\mu_{\pm,N} = \mu_N \pm \mu_{a,N} \tag{42}$$

are defined here. The graphene sheet in the $N$-th layer is assumed to be magnetically biased in the z-axis, which has the following conductivity tensor [71]:

$$\bar{\bar{\sigma}}_N\left(\omega, \mu_c(E_0), \Gamma, T, B_0\right) = \begin{pmatrix} \sigma_{O,N} & \sigma_{H,N} \\ -\sigma_{H,N} & \sigma_{O,N} \end{pmatrix} \tag{43}$$

In (43), $\omega$ is the radian frequency, $\Gamma$ is the phenomenological electron scattering rate ($\Gamma = 1/\tau$, where $\tau$ is the relaxation time), $T$ is the temperature, $\mu_C$ is the chemical potential which can be altered by chemical doping or electrostatic bias $E_0$, and $B_0$ is the applied magnetostatic bias field [71]. In addition, $\sigma_{O,N}, \sigma_{H,N}$ are the direct and indirect (Hall) conductivities of the graphene, which are obtained by using Kubo's relations reported in [71].

To analysis the structure of Fig.1, we suppose that the $N$-th layer of the anisotropic medium has the characteristics equation of

$$s^4 + A_{1,N}s^2 + A_{2,N} = 0 \tag{44}$$

Where

$$A_{1,N} = \left(\frac{\mu_N \varepsilon_N}{\mu_{\|,N} \varepsilon_{\|,N}}\right)\left( \left(k_0^2 \varepsilon_{\|,N}\mu_{\perp,N} - k_\rho^2\right)\frac{\mu_{\|,N}}{\mu_N} + \left(k_0^2 \varepsilon_{\perp,N}\mu_{\|,N} - k_\rho^2\right)\frac{\varepsilon_{\|,N}}{\varepsilon_N} + k_0^2 \mu_{\|,N}\varepsilon_{\|,N}\left(\frac{\varepsilon_{a,N}}{\varepsilon_N} + \frac{\mu_{a,N}}{\mu_N}\right)^2 \right) \tag{45}$$

$$A_{2,N} = \left(\frac{\mu_N \varepsilon_N}{\mu_{\|,N}\varepsilon_{\|,N}}\right)\left[\left(k_0^2 \varepsilon_{\|,N}\mu_{\perp,N} - k_\rho^2\right)\cdot\left(k_0^2 \varepsilon_{\perp,N}\mu_{\|,N} - k_\rho^2\right)\right] \tag{46}$$

Then, the roots of characteristics equation for each medium are expressed as,

$$k_{z,2N-1} = \sqrt{\frac{-A_{1,N} + \sqrt{A_{1,N}^2 - 4A_{2,N}}}{2}} \tag{47}$$

$$k_{z,2N} = \sqrt{\frac{-A_{1,N} - \sqrt{A_{1,N}^2 - 4A_{2,N}}}{2}} \tag{48}$$

Therefore, the roots of characteristics equations for various regions of Fig.1 can be considered

$$k_z = \begin{cases} k_{z,1}, k_{z,2} & i=1,2 \;;\; for\, layer\, N=1 \\ \ldots & \ldots \\ k_{z,2N-1}, k_{z,2N} & i=2N-1, 2N \;;\; for\, layer\, N \\ k_{z,2N+1}, k_{z,2N+2} & i=2N+1, 2N+2 \;;\; for\, layer\, N+1 \end{cases} \qquad (49)$$

In (49), *N* indicates the number of the layer and *i* shows the index of the root for that layer. Now, one should write the electromagnetic fields $H_m(z)$ and $E_m(z)$ for various regions,

$$H_m(z) = \begin{cases} A^+_{m,1,1}e^{+k_{z,1}z} + A^+_{m,2,1}e^{+k_{z,2}z} & z<0 \\ A^+_{m,3,2}e^{+k_{z,3}z} + A^+_{m,4,2}e^{+k_{z,4}z} + \\ A^-_{m,3,2}e^{-k_{z,3}z} + A^-_{m,4,2}e^{-k_{z,4}z} & 0<z<d_2 \\ \ldots \\ A^+_{m,2N+1,N+1}e^{+k_{z,2N+1}z} + A^+_{m,2N+2,N+1}e^{+k_{z,2N+2}z} + \\ A^-_{m,2N+1,N+1}e^{-k_{z,2N+1}z} + A^-_{m,2N+2,N+1}e^{-k_{z,2N+2}z} & \sum_{k=2}^{N}d_k < z < \sum_{k=2}^{N}d_k + d_S \\ B^-_{m,2N+1,N+1}e^{-k_{z,2N+1}z} + B^-_{m,2N+2,N+1}e^{-k_{z,2N+2}z} & z > \sum_{k=2}^{N}d_k + d_S \end{cases} \qquad (50)$$

$$E_m(z) = \begin{cases} T^+_{1,1}A^+_{m,1,1}e^{+k_{z,1}z} + T^+_{2,1}A^+_{m,2,1}e^{+k_{z,2}z} & z<0 \\ T^+_{3,2}A^+_{m,3,2}e^{+k_{z,3}z} + T^+_{4,2}A^+_{m,4,2}e^{+k_{z,4}z} + \\ T^-_{3,2}A^-_{m,3,2}e^{-k_{z,3}z} + T^-_{4,2}A^-_{m,4,2}e^{-k_{z,4}z} & 0<z<d_2 \\ \ldots \\ T^+_{2N+1,N+1}A^+_{m,2N+1,N+1}e^{+k_{z,2N+1}z} + T^+_{2N+2,N+1}A^+_{m,2N+2,N+1}e^{+k_{z,2N+2}z} + \\ T^-_{2N+1,N+1}A^-_{m,2N+1,N+1}e^{-k_{z,2N+1}z} + T^-_{2N+2,N+1}A^-_{m,2N+2,N+1}e^{-k_{z,2N+2}z} & \sum_{k=2}^{N}d_k < z < \sum_{k=2}^{N}d_k + d_S \\ T^-_{2N+1,N+1}B^-_{m,2N+1,N+1}e^{-k_{z,2N+1}z} + T^-_{2N+2,N+1}B^-_{m,2N+2,N+1}e^{-k_{z,2N+2}z} & z > \sum_{k=2}^{N}d_k + d_S \end{cases} \qquad (51)$$

Where

$$T^{\pm}_{i,N} = \frac{\pm 1}{k_0 \varepsilon_{\parallel,N}\left(\frac{\varepsilon_{a,N}}{\varepsilon_N} + \frac{\mu_{\alpha,N}}{\mu_N}\right)}\left(\frac{\mu_{\parallel,N}}{\mu_N}k_{z,i} + \frac{1}{k_{z,i}}\left(k_0^2 \varepsilon_{\perp,N}\mu_{\parallel,N} - k_\rho^2\right)\right) \qquad i=2N-1, 2N, \; N=1,2,3,\ldots \qquad (52)$$

are defined similarly to the relation (30). The transverse components of electric and magnetic fields are expressed:

$$\begin{pmatrix} E^{\pm}_{r,i} \\ H^{\pm}_{r,i} \end{pmatrix} = \bar{\bar{Q}}^{Pos,\pm}_{i,N}\frac{\partial}{\partial r}\begin{pmatrix} E^{\pm}_{z,i} \\ H^{\pm}_{z,i} \end{pmatrix} + \frac{m}{r}\bar{\bar{Q}}^{Neg,\pm}_{i,N}\begin{pmatrix} E^{\pm}_{z,i} \\ H^{\pm}_{z,i} \end{pmatrix} \qquad i=2N-1, 2N \qquad (53)$$

$$j\begin{pmatrix} E_{\varphi,i}^{\pm} \\ H_{\varphi,i}^{\pm} \end{pmatrix} = \bar{\bar{Q}}_{i,N}^{Neg,\pm} \frac{\partial}{\partial r}\begin{pmatrix} E_{z,i}^{\pm} \\ H_{z,i}^{\pm} \end{pmatrix} + \frac{m}{r}\bar{\bar{Q}}_{i,N}^{Pos,\pm}\begin{pmatrix} E_{z,i}^{\pm} \\ H_{z,i}^{\pm} \end{pmatrix} \qquad (54)$$

Where the Q-matrices in (53), (54) are defined as follows:

$$\bar{\bar{Q}}_{i,N}^{Pos,\pm} = \frac{1}{2}\left[\frac{1}{k_{z,i}^2 + k_0^2 \varepsilon_{+,N}\mu_{+,N}}\begin{pmatrix} \pm k_{z,i} & -\omega\mu_0\mu_{+,N} \\ \omega\varepsilon_0\varepsilon_{+,N} & \pm k_{z,i} \end{pmatrix} + \frac{1}{k_{z,i}^2 + k_0^2\varepsilon_{-,N}\mu_{-,N}}\begin{pmatrix} \pm k_{z,i} & \omega\mu_0\mu_{-,N} \\ -\omega\varepsilon_0\varepsilon_{-,N} & \pm k_{z,i} \end{pmatrix}\right]$$
(55)

$$\bar{\bar{Q}}_{i,N}^{Neg,\pm} = \frac{1}{2}\left[\frac{1}{k_{z,i}^2 + k_0^2 \varepsilon_{+,N}\mu_{+,N}}\begin{pmatrix} \pm k_{z,i} & -\omega\mu_0\mu_{+,N} \\ \omega\varepsilon_0\varepsilon_{+,N} & \pm k_{z,i} \end{pmatrix} - \frac{1}{k_{z,i}^2 + k_0^2\varepsilon_{-,N}\mu_{-,N}}\begin{pmatrix} \pm k_{z,i} & \omega\mu_0\mu_{-,N} \\ -\omega\varepsilon_0\varepsilon_{-,N} & \pm k_{z,i} \end{pmatrix}\right]$$
(56)

Now, the characteristic equation is achieved by applying boundary conditions. For the graphene sheet sandwiched between two magnetic materials, one can write the following boundary conditions in general form:

$$E_{r,N} = E_{r,N+1}, E_{\varphi,N} = E_{\varphi,N+1} \qquad N = 1,2,3,.... \qquad (57)$$

$$H_{r,N+1} - H_{r,N} = -\sigma_{H,N} E_{r,N} + \sigma_{O,N} E_{\varphi,N}, H_{\varphi,N+1} - H_{\varphi,N} = -\left(\sigma_{O,N} E_{r,N} + \sigma_{H,N} E_{\varphi,N}\right) \qquad N = 1,2,3,....$$
(58)

And for the last boundary at $z = \sum_{k=2}^{N} d_k + d_s$,

$$E_{r,N+1}^{>}\Big|_{z=\sum_{k=2}^{N} d_k + d_S} - E_{r,N+1}^{<}\Big|_{z=\sum_{k=2}^{N} d_k + d_S} = -M_{s\varphi}, \ E_{\varphi,N+1}^{>}\Big|_{z=\sum_{k=2}^{N} d_k + d_S} - E_{\varphi,N+1}^{<}\Big|_{z=\sum_{k=2}^{N} d_k + d_S} = M_{sr} \qquad (59)$$

$$H_{r,N+1}^{>}\Big|_{z=\sum_{k=2}^{N} d_k + d_S} - H_{r,N+1}^{<}\Big|_{z=\sum_{k=2}^{N} d_k + d_S} = J_{sr}, \ H_{\varphi,N+1}^{>}\Big|_{z=\sum_{k=2}^{N} d_k + d_S} - H_{\varphi,N+1}^{<}\Big|_{z=\sum_{k=2}^{N} d_k + d_S} = -J_{s\varphi} \qquad (60)$$

In (59)-(60), $M_{sr}, M_{s\varphi}, J_{sr}, J_{s\varphi}$ are $r$ and $\varphi$-components of magnetic and electric currents at $z = \sum_{k=2}^{N} d_k + d_s$, respectively. Now, by applying the boundary conditions expressed in (57)-(60), the final matrix representation is derived:

$$\bar{\bar{S}}_{4N+4,4N+4} \cdot \begin{pmatrix} A_{m,1,1}^{+} \\ A_{m,2,1}^{+} \\ A_{m,3,2}^{+} \\ A_{m,4,2}^{+} \\ \vdots \\ A_{m,2N+1,N+1}^{-} \\ A_{m,2N+2,N+1}^{-} \\ B_{m,2N+1,N+1}^{-} \\ B_{m,2N+2,N+1}^{-} \end{pmatrix}_{4N+4,1} = \begin{pmatrix} 0 \\ 0 \\ 0 \\ 0 \\ \vdots \\ -M_{s\varphi} \\ M_{sr} \\ J_{sr} \\ -J_{s\varphi} \end{pmatrix}_{4N+4,1} \qquad (61)$$

In (61), the matrix $\bar{\bar{S}}$ is

$$\bar{\bar{S}} = \begin{pmatrix}
P^+_{1,1,1} & P^+_{2,1,1} & -P^+_{3,2,1}e^{+k_{z,3}d_2} & \ldots & \ldots & 0 & 0 & 0 & & 0 \\
R^+_{1,1,1} & R^+_{2,1,1} & -R^+_{3,2,1}e^{+k_{z,3}d_2} & \ldots & \ldots & 0 & 0 & 0 & & 0 \\
\ldots & \ldots & \ldots & & \ldots & 0 & 0 & 0 & & 0 \\
\ldots & \ldots & \ldots & & \ldots & 0 & 0 & 0 & & 0 \\
\ldots & \ldots & \ldots & \ldots & \ldots & \ldots & \ldots & \ldots & & \ldots \\
0 & 0 & 0 & 0 & \ldots & \ldots & \ldots & \ldots & & \ldots \\
0 & 0 & 0 & 0 & \ldots & \ldots & \ldots & \ldots & & \ldots \\
0 & 0 & 0 & 0 & \ldots & \ldots & \ldots & \ldots & & \ldots \\
0 & 0 & 0 & 0 & \ldots & \ldots & \ldots & \ldots & R^-_{2N+2,N+1,2}e^{-k_{z,2N+2}\left(\sum_{k=2}^{N}d_k+d_S\right)}
\end{pmatrix} \quad (62)$$

Where the following relations have been used in (62),

$$\begin{pmatrix} P^{\pm}_{i,N,1}(r) \\ P^{\pm}_{i,N,2}(r) \end{pmatrix} = \left[ \bar{\bar{Q}}^{Pos,\pm}_{i,N} k_\rho J'_m(k_\rho r) + \frac{m}{r}\bar{\bar{Q}}^{Neg,\pm}_{i,N} J_m(k_\rho r) \right] \begin{pmatrix} T^{\pm}_{i,N} \\ 1 \end{pmatrix} \quad i=2N-1, 2N \quad (63)$$

$$\begin{pmatrix} R^{\pm}_{i,N,1}(r) \\ R^{\pm}_{i,N,2}(r) \end{pmatrix} = -j\left[ \bar{\bar{Q}}^{Neg,\pm}_{i,N} k_\rho J'_m(k_\rho r) + \frac{m}{r}\bar{\bar{Q}}^{Pos,\pm}_{i,N} J_m(k_\rho r) \right] \begin{pmatrix} T^{\pm}_{i,N} \\ 1 \end{pmatrix} \quad i=2N-1, 2N \quad (64)$$

$N = 1, 2, 3, \ldots$

Now, our analytical model has been completed for the general multi-layer structure. It should be emphasized that the matrix $\bar{\bar{S}}$ is an important matrix because its determinant (set $det(\bar{\bar{S}}) = 0$) obtains the dispersion relation (or the propagation constant) of the general multi-layered structures. By obtaining the propagation constant, the plasmonic parameters of the general multi-layer structure such as the effective index ($n_{eff} = Re(k_\rho/k_0)$), the propagation losses ($\alpha = Im(k_\rho/k_0)$), the decay length ($l_z = 1/|Im(k_z)|$), and figure of merit based on the quality factor (or briefly called "benefit-to-cost ratio", $FOM = Re(k_\rho)/2\pi Im(k_\rho)$) [72] are achieved easily.

It is worth to be mentioned that both the electric and magnetic currents have been considered as excitation sources (located at $z = d_S + \sum_{k=2}^{N} d_k$) to generalize our analysis. To excite magneto-plasmons, one of these components ($M_{sr}, M_{s\varphi}, J_{sr}, J_{s\varphi}$) is sufficient. For instance, the excitation source can be assumed a magnetic ring source with the radius of $\acute{r}$ ( $\bm{M} = M_{s\varphi}\hat{\bm{a}}_\varphi = \frac{1}{\acute{r}}\delta[r - \acute{r}]\delta[z - (d_S + \sum_{k=2}^{N} d_k)]\hat{\bm{a}}_\varphi$, where $\delta$ indicates "delta function"), located symmetrically about the z-axis at $z = d_S + \sum_{k=2}^{N} d_k$. Since the problem is symmetric in φ, we must have $m = 0$ ($\frac{\partial}{\partial \varphi} = 0$) for this excitation source. It should be noted that we have utilized the modal analysis (Eigen-value and Eigen-frequency analysis) in our simulations.

The presented model does not consider the "direct magnetic coupling of the neighboring magnetic layers" and the "magnetic anisotropic effects" arising when the magnetic layers are thin. We believe that the analytical model works properly when the thicknesses of the magnetic layers are in the range of microns down to several ten nanometers. In what follows, we will consider four exemplary structural variants to show the richness of the proposed general structure regarding the related specific plasmonic wave phenomena and effects.

## 4. Special Cases of the General Structure: Results and Discussions

This section studies the modal properties of four exemplary structural variants. The first and second waveguides have been chosen to verify the performance and high accuracy of the theoretical model outlined in the previous section.

The first example studies a graphene sheet, with an anisotropic conductivity tensor of $\bar{\bar{\sigma}}$, sandwiched between two dielectric layers with different permittivities. The graphene sheet has been biased magnetically in this structure, which supports hybrid TE-TM plasmonic mode. The second example is a three-layer structure, constituting Air-Graphene-$SiO_2$-Si layers. In this structure, the graphene layer has been deposited on $SiO_2$-Si layers and its conductivity is characterized by isotropic surface conductivity. This waveguide has been studied to confirm the validity of the proposed model, especially when the graphene layer is isotropic.

To show the richness of the proposed general structure regarding the related specific plasmonic wave phenomena and effects, two novel graphene-based structures have been suggested and investigated as the third and the fourth structures. The third example introduces a novel ferrite-graphene waveguide, which has a non-reciprocity effect. This waveguide is able to control the plasmonic features via the chemical potential and magnetic bias. As a fourth example, a planar graphene-based structure with a gyro-electric substrate has been considered. The gyroelectric layer has an anisotropic permittivity, which produces a non-reciprocity effect. All of these structures have been simulated in the software and the full-wave simulation results are compared with theoretical ones to show the validity of the analytical model.

### 4.1. The First Structure: The Anisotropic Graphene Sheet Sandwiched Between two Dielectric Layers

As a first example, a familiar structure has been studied. Fig. 2 illustrates this configuration, where the anisotropic graphene sheet has been located between two different dielectric layers ($\mu_1 = \mu_2 = \mu_0$) with permittivities of $\varepsilon_1$, $\varepsilon_2$. The graphene sheet has been biased magnetically in the z-direction ($\boldsymbol{B_0} = B_0 \hat{a}_z$). Therefore, the graphene is modeled by a conductivity tensor given in (43). As mentioned before, the elements of this tensor are obtained by Kubo's formula [71]. The studied case is an important structure because it supports hybrid TE-TM plasmonic mode and the plasmonic features can be adjusted via the magnetic bias and the chemical doping.

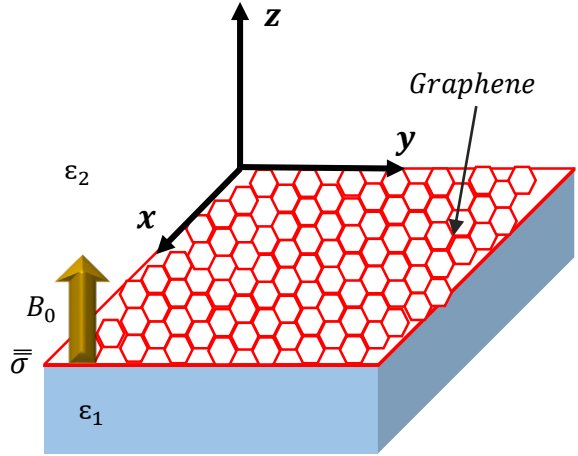

**Fig. 2.** The graphene sheet sandwiched between two different dielectric layers.

By applying the theoretical relations outlined in the previous section and doing some rigorous mathematical procedures, the dispersion equation is derived:

$$\left( \frac{\sqrt{k_0^2 \varepsilon_1 - k_\rho^2} + \sqrt{k_0^2 \varepsilon_1 - k_\rho^2}}{\omega \mu_0} + \sigma_o \right) \times \left( \frac{\omega \varepsilon_1}{\sqrt{k_0^2 \varepsilon_1 - k_\rho^2}} + \frac{\omega \varepsilon_2}{\sqrt{k_0^2 \varepsilon_2 - k_\rho^2}} + \sigma_o \right) = \sigma_H^2 \qquad (65)$$

The above equation can be simplified to the following relation for the case $k_\rho \gg k_0$:

$$k_\rho \approx \frac{j k_0}{4 \eta_0 \sigma_o} \left[ \left( 2(\varepsilon_1 + \varepsilon_2) + \eta_0^2 (\sigma_o^2 + \sigma_H^2) \right) + \sqrt{\left[ 2(\varepsilon_1 + \varepsilon_2) + \eta_0^2 (\sigma_o^2 + \sigma_H^2) \right]^2 - 8 \eta_0^2 (\sigma_o^2 + \sigma_H^2)} \right] \qquad (66)$$

Where $\eta_0$ is the impedance of the free space. For simplicity, we suppose that the upper dielectric is air ($\varepsilon_2 = \varepsilon_0$) and the bottom layer is assumed to be Si ($\varepsilon_1 = 11.9\,\varepsilon_0$), which is common for practical applications. The graphene parameters are $T = 300\,K, \tau = 0.14\,ps$.

Firstly, we compare the theoretical and full-wave simulation results to validate the analytical model outlined in the previous section. Fig. 3 demonstrates the theoretical and simulation results of the effective index and FOM for the various external magnetic fields ($B_0 = 0,1,4\,T$). In this figure, the chemical potential of graphene is supposed to be 0.12 eV. A full agreement between the simulation and theoretical results is observed, which confirms the high accuracy and the good performance of our analytical model. The high accuracy of our analytical model in computing the plasmonic characteristics of the waveguide arises from this matter that no approximations have been used in deriving these complicated relations. The hybrid plasmonic wave excited in the waveguide appears only for $Im[\sigma_0] > 0$, as expected. One can see that a very high effective index of about 3900 is obtained at the frequency of 32 THz for the magnetic bias B= 4T. The results indicate that the increment of the magnetic bias enhances the field confinement, which is very desirable in nano-technologies.

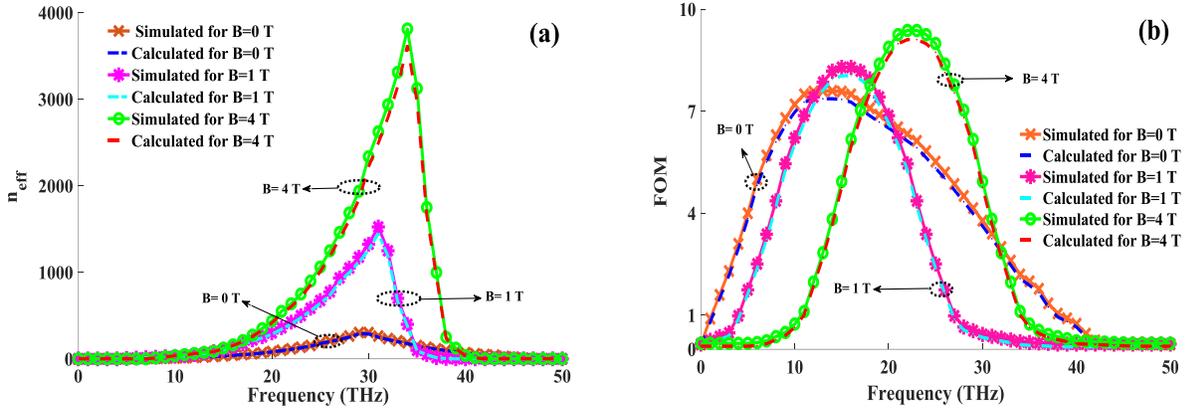

**Fig. 3.** The theoretical and simulation results of the effective index and FOM as a function of frequency for the various magnetic fields ($B_0 = 0,1,4\,T$). The chemical potential is supposed to be 0.12 eV.

To study the influence of the chemical potential on the plasmonic characteristics, the effective index and FOM have been depicted for the chemical potential of 0.3 eV in Fig. 4. It is clear that the effect of the chemical potential is dominated over the external magnetic field in this case. As mentioned before, the main property of the general multi-layered structure (shown in Fig.1) is its ability to control the plasmonic features such as the effective index via the chemical doping and magnetic bias field. To see the effect of the magnetic bias on the modal features of the plasmonic waveguide, the structure should be used only for low chemical potentials (~0.12 eV).

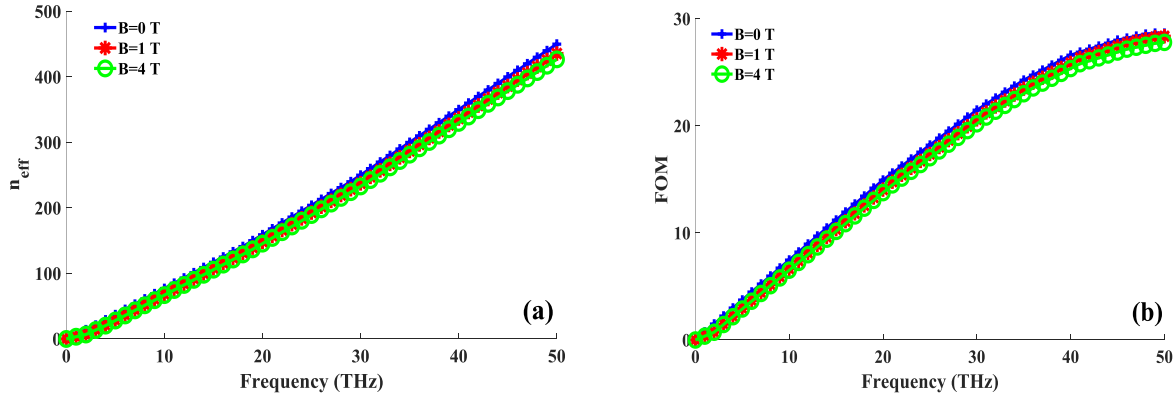

**Fig. 4.** The theoretical results of the effective index and FOM for the various external magnetic fields ($B_0 = 0,1,4\,T$). The chemical potential is assumed to be 0.3 eV.

As a final point for this sub-section, the decay length has been represented in Fig. 5 for various values of chemical potentials and magnetic fields. Again, it is observed that the influence of the magnetic bias is limited for high chemical potentials. In addition, as the chemical potential increases, the decay length shows a blue-shift. The theoretical model could estimate the propagation features of the first example (Fig. 2) as well. In the next sub-section, another useful structure will be studied in detail.

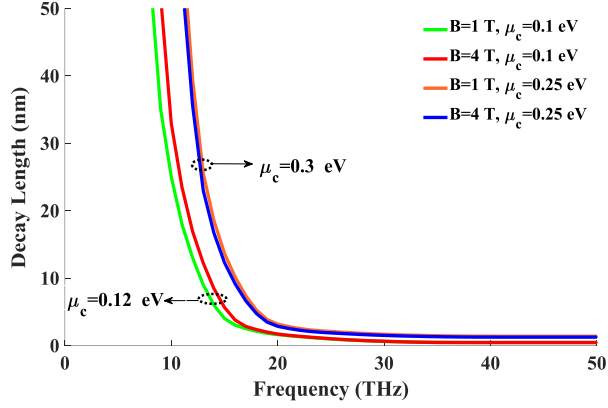

**Fig. 5.** The Decay length versus the frequency for the various values of magnetic fields and chemical potentials.

*4.2. The Second Structure: The Graphene-Based Three-layered Structure*

Fig. 6 illustrates the general configuration of the three-layer structure. The graphene sheet has an isotropic surface conductivity of σ, which means that only the diagonal elements of (43) exist here (the graphene has been biased via electrostatic field or chemical doping). Therefore, no hybrid mode is excited in this structure and TE and TM modes propagate separately. The structure has been chosen to indicate that our theoretical expressions also work for simple configurations such as the isotropic graphene sheets with dielectric layers.

The dispersion relations for TM and TE modes of this waveguide are found by applying the analytical expressions of the previous section. Here, we investigate only TM mode due to its good mode confinement. The dispersion relation for TM mode are derived:

$$\left[\frac{1+C_{TM}}{1-C_{TM}}\right] \times \left[\frac{1+D_{TM}}{1-D_{TM}}\right] = \exp\left(-2\sqrt{k_0^2 \varepsilon_2 - k_\rho^2}\, d\right) \tag{67}$$

Where the coefficients of

$$C_{TM} = \frac{\sqrt{k_0^2 \varepsilon_2 - k_\rho^2}}{\sqrt{k_0^2 \varepsilon_3 - k_\rho^2}} \times \frac{\varepsilon_3}{\varepsilon_2} \times \left(1 + j\frac{\sigma \sqrt{k_0^2 \varepsilon_3 - k_\rho^2}}{\omega \varepsilon_0 \varepsilon_3}\right) \tag{68}$$

$$D_{TM} = \frac{\sqrt{k_0^2 \varepsilon_2 - k_\rho^2}}{\sqrt{k_0^2 \varepsilon_1 - k_\rho^2}} \times \frac{\varepsilon_1}{\varepsilon_2} \tag{69}$$

have been used in the relation (67).

For simplicity and without loss of the generality, we assume that the third region is air ($\varepsilon_3 = \varepsilon_0$) and graphene sheet has been deposited on the SiO$_2$-Si substrate ($\varepsilon_1 = 11.9\,\varepsilon_0$, $\varepsilon_2 = 2.09\,\varepsilon_0$) with a thickness of $d = 5\,\mu m$. In practical design, the Si- substrate is applied to change the Fermi level of graphene, which is known as "back-gate" in the literature. Here, the graphene parameters are $\mu_c = 0.23\,eV, \tau = 0.5\,ps$.

The theoretical and simulation results of the effective index and the propagation loss for TM mode has been shown in Fig. 7. A full agreement between the simulation and analytical results is seen. As observed in this figure, the effective index increases as the frequency increases. This matter happens due to the reduction of the conductivity for the frequency increment. Besides, the propagation loss is affected by two opposite trends. Firstly, as the frequency increases, the conductivity reduces, which results in the losses increment. But on the other hand, as the wavelength decreases, the penetration of electromagnetic fields inside the Si layer decreases and thus the loss reduces. These opposite trends occur for the propagation loss diagram.

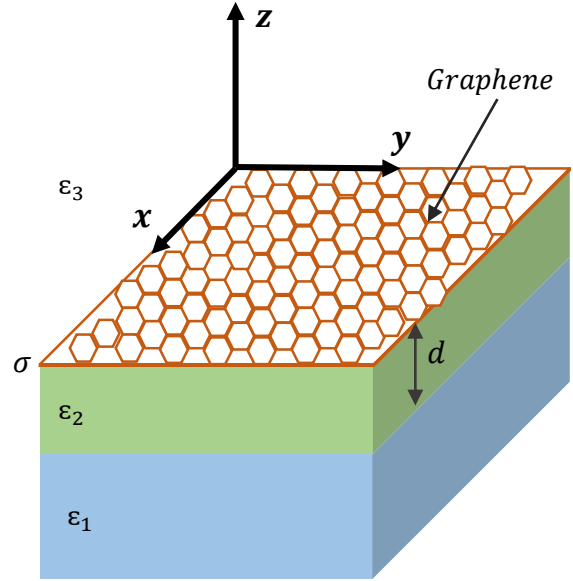

**Fig. 6.** The graphene-based three-layer structure

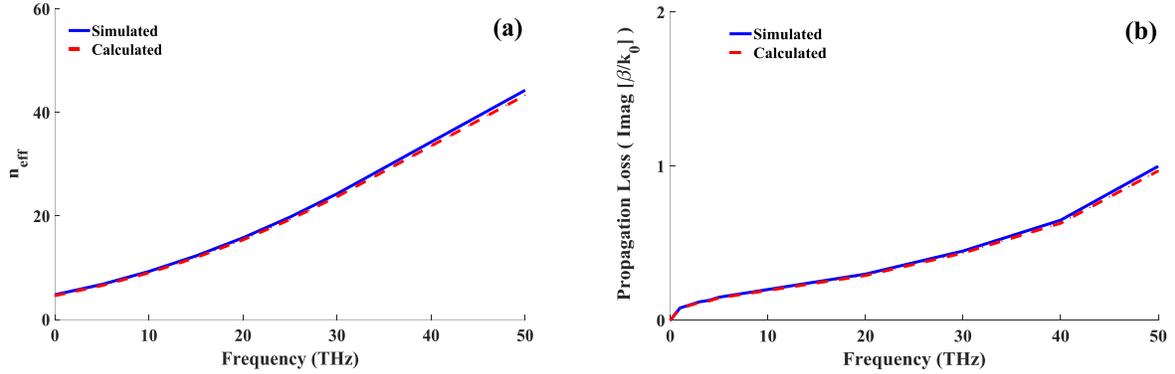

**Fig. 7.** Theoretical and simulation results for the effective index and the propagation loss of the Air-Graphene-SiO$_2$-Si structure for TM mode.

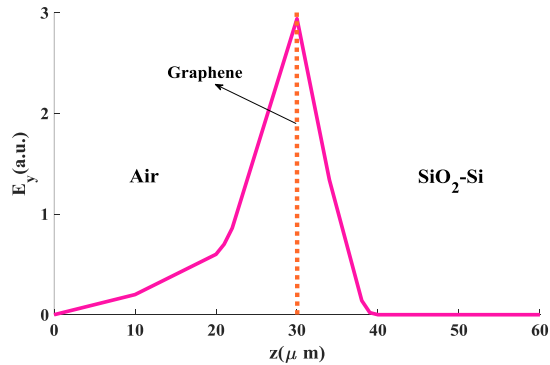

**Fig. 8.** Theoretical result of $E_y$ component for TM mode of the Air-Graphene- SiO$_2$-Si structure at the frequency 10 THz.

The $E_y$ distributions for TM mode of the Air-Graphene- SiO$_2$-Si structure has been depicted at the frequency of 10 THz. This figure reveals that the peak of the electric field is at the graphene boundary and the fields reduce exponentially in other layers, which indicates the field confinement at the graphene sheet. The theoretical results showed that our analytical expressions are very useful and accurate for the second structure. In the third example, we will study the hybrid plasmonic mode in the novel configuration of graphene-ferrite at low THz frequencies (0.1-0.25 THz).

*4.3. The Third Structure: Hybrid Ferrite-Graphene Waveguide*

As a third example, a novel graphene-based structure has been introduced and studied, in which the hybridization of graphene-ferrite at low THz frequencies can control non-reciprocal magneto-plasmons via chemical doping and magnetic bias. As mentioned before, the usage of the ferrite materials in low THz frequencies has been reported by some research articles [56-59]. Hence, it is common to use ferrite materials for designing novel plasmonic devices at low THz frequencies.

The schematic of the proposed waveguide has been shown in Fig. 9, where the graphene layers have been placed on the top and the bottom surfaces of the anisotropic layer. For simplicity, we assume that the dielectric layer at top of the structure is air ($\varepsilon_4 = \varepsilon_0, \mu_4 = \mu_0$). In Fig. 9, the anisotropic layer is supposed to be a ferrite with the thickness of $t$ and it is located on SiO$_2$-Si layers ($\varepsilon_1 = 11.9\,\varepsilon_0$, $\varepsilon_2 = 2.09\,\varepsilon_0$, $\mu_1 = \mu_2 = \mu_0$). The geometrical parameters are $d = 2\,\mu m, t = 9\,\mu m$. The magnetic bias of $B_0 = 4\,T$ has been applied in the z-direction. In the simulations, the highly doped graphene layers have been used ($\mu_c = 0.9\,eV, \tau = 0.96\,ps, T = 300\,K$). The ferrite material is chosen Lu$_{2.1}$Bi$_{0.9}$I$_5$G$_{12}$ (or briefly called "LuBiIG") with the gyromagnetic ratio of 1.758×10$^{11}$rad/T.s, the magnetic line width of 5.1 Oe, the saturation magnetization of 1560 G, and the relative permittivity of 4.85 ($\varepsilon_3 = 4.85\,\varepsilon_0$).

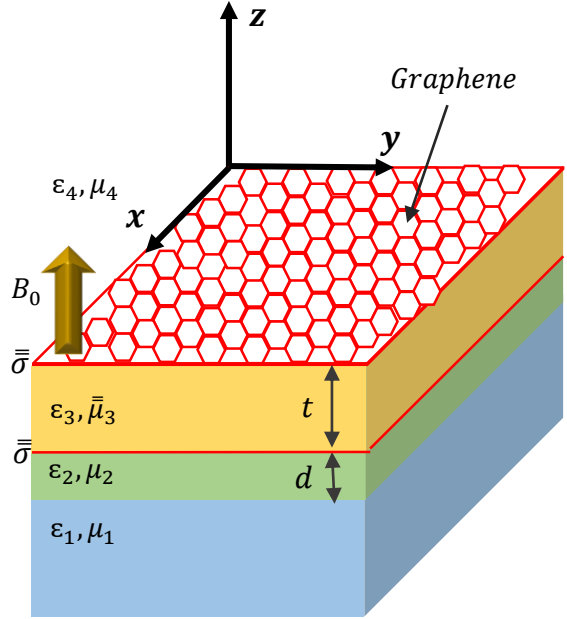

**Fig. 9.** The schematic of the hybrid ferrite–graphene structure.

The dispersion relation, the distributions of the electromagnetic fields and the plasmonic parameters such as the effective index can be derived for this structure by utilizing the analytical model outlined in section 3. Here, we do not mention and report these equations due to their complicated mathematical relations.

Fig. 10 demonstrates the simulation and theoretical results for the effective index and the propagation loss. A very good agreement between the simulation and theoretical results is observed, which validates our analytical model. It can be seen that a nonreciprocal resonance occurs in the vicinity of the Larmor frequency in Fig. 10. We believe that this non-reciprocal behavior at low THz frequencies is caused by the permeability tensor of the ferrite layer, but the graphene conductivity can change the amplitude amount of this resonance. One of the popular ways to study the non-reciprocity effect is reversing the direction of the external magnetic bias. Consequently, the magnetic bias has been reversed in the (–z)-direction to show the non-reciprocity effect, as seen in Fig. 10. For B = 4 T, the maximum effective index reaches to $n_{eff} = 16$ at frequency 0.13 THz.

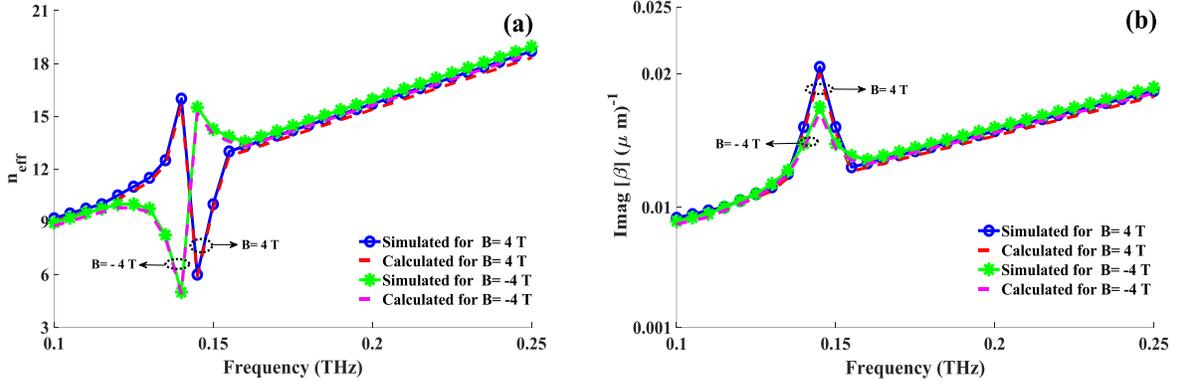

**Fig. 10.** Simulation and theoretical results of the effective index and the propagation loss for various values of magnetic fields. The chemical potential of the graphene is 0.9 eV.

The effective index is one of the main parameters for designing the plasmonic devices, which shows the mode confinement in the plasmonic structure. Accordingly, we investigate the influence of controlling mechanisms such as chemical doping on the effective index. Due to the high accuracy of our proposed model, only the analytical results of the effective indices have been depicted in Fig. 11. A blue-shift is observed as the external magnetic field increases. This is an important matter for designing the graphene-ferrite waveguide to work in the specific frequency by choosing the appropriate magnetic bias. For instance, it can be estimated that the magnetic bias required for the proposed structure to work at 35 GHz is about 1T.

Finally, we study the influence of chemical potential on the effective index in the absence of magnetic bias (B = 0 T). This figure reveals that the propagation wavelength of the plasmonic mode increases for high chemical potentials. Indeed, as the chemical potential of the graphene layer increases, the effective index (or the real part of the propagation constant) decreases. As a result, the propagation wavelength of SPPs increases ($\lambda_{SPP} = 2\pi/Re(\beta)$).

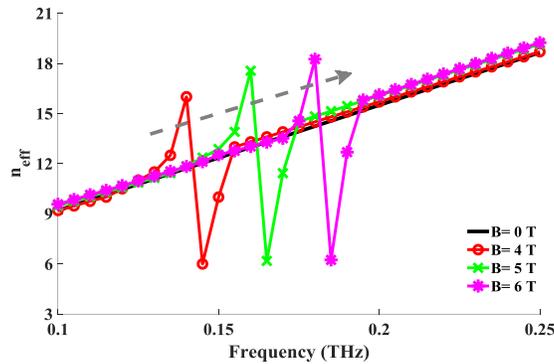

**Fig. 11.** Analytical results of the effective index for the graphene-ferrite waveguide for various values of magnetic fields. The chemical potential of the graphene is 0.9 eV.

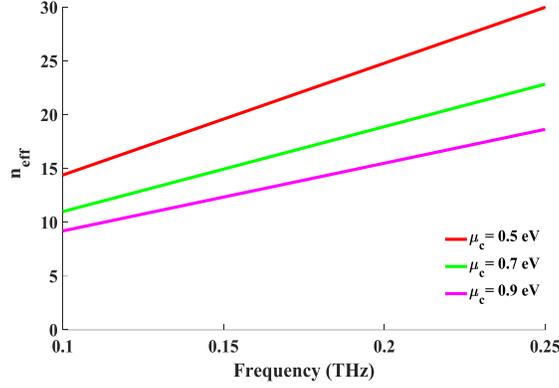

**Fig. 12.** The effective index of the hybrid graphene-ferrite waveguide as a function of the frequency for various values of chemical potentials, in the absence of magnetic bias (B = 0 T).

### 4.4. The Fourth Structure: The Graphene-Based Waveguide with Gyro-electric Substrate

In the last example, a novel hybridization of the anisotropic graphene layer and the gyro-electric material has been introduced and investigated. The schematic of the proposed structure has been represented in Fig. 13, where the gyro-electric material has been sandwiched between two graphene layers. The external magnetic field is applied in the z-axis, and thus the conductivity of the graphene has a tensor form. As in similar structures studied in previous sub-sections, the hybrid SPP wave can be controlled by altering the external magnetic field and chemical doping.

The dispersion relation for this waveguide is calculated by using the matrix representation defined in (61)-(62). Due to the complicated form of mathematical equations for this waveguide, they are not reported here. To simulate the proposed structure, the gyroelectric substrate is assumed to be n-type InSb with the thickness of $s = 50\ nm$, which its parameters are $\mu_2 = \mu_0, \varepsilon_\infty = 15.68,\ m^* = 0.022 m_e, n_s = 1.07 \times 10^{17}/cm^3, v = 0.314 \times 10^{13} s^{-1}$ and $m_e$ is the electron's mass. The upper dielectric is air ( $\varepsilon_3 = \varepsilon_0, \mu_3 = \mu_0$) and the background substrate has the permittivity of 3.8 ( $\varepsilon_1 = 3.8\ \varepsilon_0, \mu_1 = \mu_0$). The graphene parameters for this structure are assumed to be $\mu_c = 0.45\ eV, \tau = 0.34\ ps, T = 300\ K$.

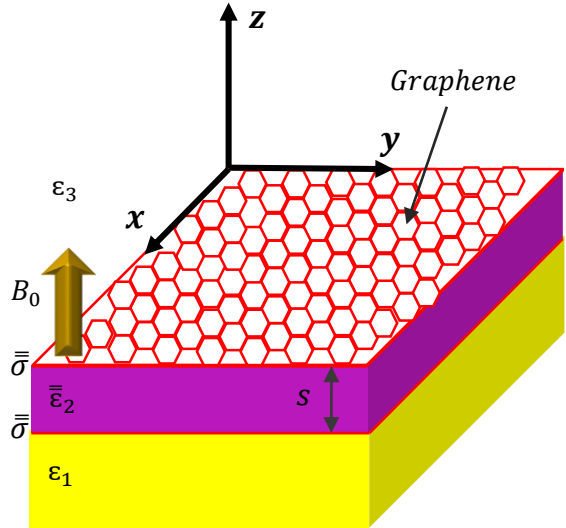

**Fig. 13.** Plasmonic graphene-based waveguide with the gyroelectric substrate

The effective indices and the propagation losses versus frequency have been shown in Fig. 14 for various values of magnetic fields. As seen in Fig. 14(a), the validity of our analytical model is confirmed. One way to check the nonreciprocity effect of the structure is reversing the direction of the DC bias magnetic field. Fig. 14 indicates that the proposed waveguide is nonreciprocal because its propagation features change as the direction of magnetic bias is reversed.

As the frequency increases and keeps out of the cyclotron frequency ($\omega \gg \omega_c$), the dependence of the propagation characteristics on the magnetic bias vanishes. This happens because the non-diagonal elements in the permittivity tensor become smaller than the diagonal elements (see relation (4)) for high frequencies. One can see from Fig. 14 (a) that the nonreciprocity resonances appear in the frequency range of 5-9 THz. In this region, applying the external DC bias field increases the absorption loss of the mode. For a better representation of the nonreciprocal resonances, the frequency region of 5-9 THz has been magnified in Fig. 14 (a).

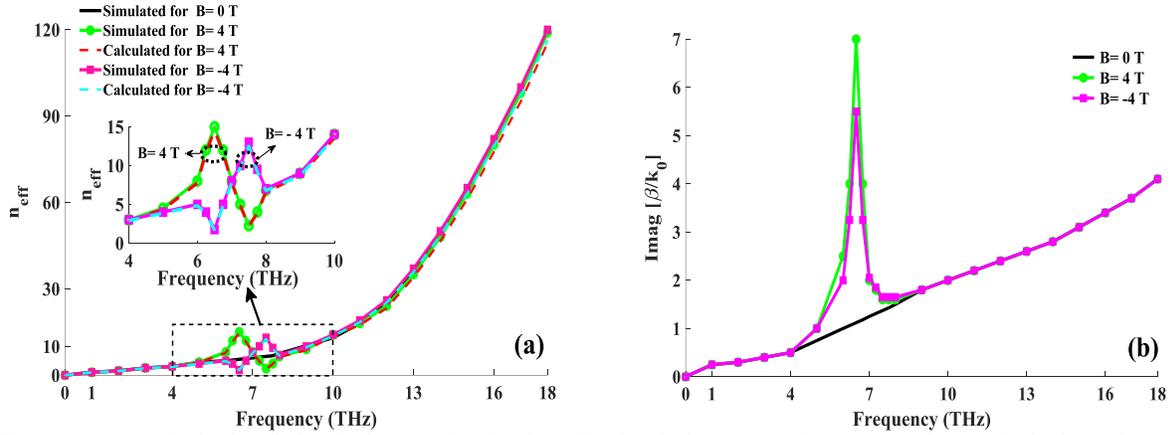

**Fig. 14.** **(a)** Analytical and simulation results for the effective index versus frequency, **(b)** Analytical results of the propagation losses for various external magnetic fields (*B= -4,0,4 T*). The chemical potential of the graphene layers are supposed to be $\mu_c = 0.45\ eV$.

FOM is the main parameter for studying the performance of the plasmonic structures. To get exact insight for the influence of the magnetic bias on plasmonic characteristics, FOM has been demonstrated for various values of magnetic fields in Fig. 15(a). This figure reveals that FOM strongly depends on the frequency in the non-reciprocity range (5-9 THz region). Indeed, non-diagonal elements play the main role in this range. These elements have a significant amplitude comparing to the diagonal elements, which means that the propagation parameters (such as the effective index and the propagation loss) strongly depend on the external magnetic bias. Therefore, changing the external magnetic field causes FOM variations (valley appears) in this range. While for the high or small frequencies, where the nonreciprocity vanishes, FOM does not depend on the magnetic field variations. Another interesting hint in Fig. 15(a) is the existence of a valley for frequencies *6.5, 7 THz*. This valley clearly indicates that FOM has a minimum point, which should not be designed at this point.

One of the important aspects of the proposed structure is its tunability by altering the chemical potential. FOM has been depicted as a function of the chemical potential in Fig. 15 (b). As seen in this figure, a minimum value exists for the FOM diagram at high frequencies. As the frequency decreases, this minimum point disappears.

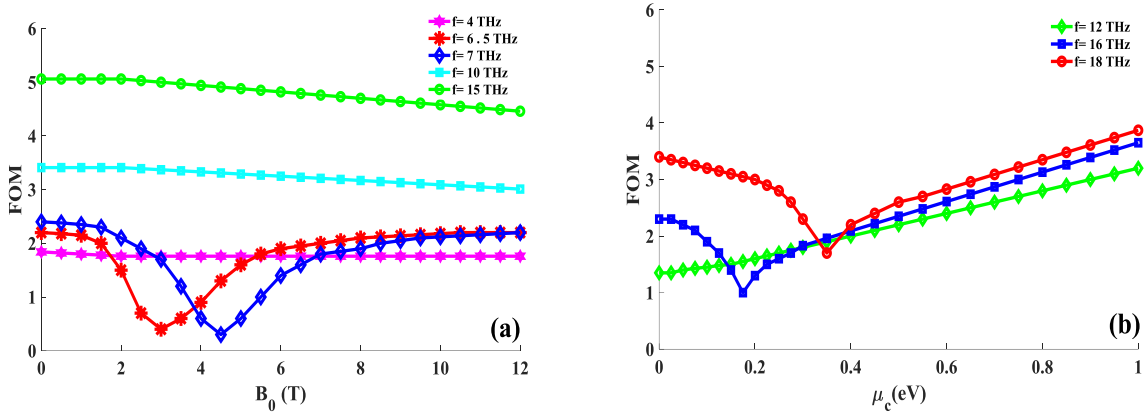

**Fig. 15. (a)** FOM as a function of magnetic bias at various frequencies (the chemical potential of the graphene layers are supposed $\mu_c = 0.45\ eV$), **(b)** FOM as a function of chemical potential at various frequencies (the applied magnetic bias is $B= 4\ T$).

## 5. Conclusion

In this article, a general theoretical model has been proposed for multi-layer structures containing anisotropic graphene sheets. This general and rich structure supports magneto-plasmons, with adjustable modal properties by varying the chemical doping and magnetic bias. Four of its exemplary structural variants have been chosen to show, first, plasmonic properties of these structures, and second, the performance and the accuracy of the proposed model. The first structure was composed of an anisotropic graphene layer sandwiched between two dielectric layers. A large value of the refractive index, amounting to 3600 e.g., at the frequency of 32 THz, was obtained for it. Then, a basic multi-layer structure in graphene plasmonics, constituting Air-Graphene-$SiO_2$-Si layers was studied to confirm the validity of the proposed model, especially when the graphene layer is isotropic. In the third example, a hybrid ferrite-graphene structure with DC magnetic bias (in the z-direction) has been considered, which supports non-reciprocal plasmonic mode. This structure is a tunable waveguide, which can control the propagation features of magneto-plasmons via chemical doping and the external magnetic field. As the fourth example, a planar graphene-based structure with the gyro-electric substrate was introduced. The gyroelectric layer had the anisotropic permittivity, which could produce the non-reciprocity effect, similar to the third example. The proposed model can be applied to design tunable graphene-based components in THz region such as absorbers, transparent electrodes used in solar cells, and hyperbolic metamaterial-based devices.